\documentclass[]{spie}  %>>> use for US letter paper
\usepackage{amsmath,amsfonts,amssymb}
\usepackage{graphicx}
\usepackage[colorlinks=true, allcolors=blue]{hyperref}

\usepackage{siunitx}
\usepackage[T1]{fontenc} % For extra glyphs (accents, etc)
\usepackage[utf8]{inputenc}

\title{Resolution and accuracy of non-linear regression of PSF with artificial neural networks}

\author[a]{Matthias Lehmann}
\author[a]{Christian Wittpahl}
\author[a]{Hatem Ben Zakour}
\author[a]{Alexander Braun}
\affil[a]{Düsseldorf University of Applied Sciences, Münsterstraße 156, 40476 Düsseldorf, Germany}

\authorinfo{Further author information: (Send correspondence to Alexander Braun)\\Alexander Braun: E-mail: alexander.braun@hs-duesseldorf.de, Telephone: +49 211 4351 3140}

\begin{document} 
\maketitle

\begin{abstract}
In a previous work we have demonstrated a novel numerical model for the point spread function (PSF) of an optical system that can efficiently model both experimental measurements and lens design simulations of the PSF. The novelty lies in the portability and the parameterization of this model, which allows for completely new ways to validate optical systems, which is especially interesting for mass production optics like in the automotive industry, but also for ophtalmology.
%allows to apply this model in basically any conceivable optical simulation scenario, 
The numerical basis for this model is a non-linear regression of the PSF with an artificial neural network (ANN). In this work we examine two important aspects of this model: the spatial resolution and the accuracy of the model. Measurement and simulation of a PSF can have a much higher resolution then the typical pixel size used in current camera sensors, especially those for the automotive industry. We discuss the influence this has on on the topology of the ANN and the final application where the modeled PSF is actually used. Another important influence on the accuracy of the trained ANN is the error metric which is used during training. The PSF is a distinctly non-linear function, which varies strongly over field and defocus, but nonetheless exhibits strong symmetries and spatial relations. Therefore we examine different distance and similarity measures and discuss its influence on the modeling performance of the ANN.
\end{abstract}


%%%%%%%%%%%%%%%%%%%%%%%%%%%%%%%%%%%%%%%%%%%%%%%%%%%%%%%%%%%%%%%%%%%%%%%%%%%%%%
\section{Introduction}
\begin{figure}[ht]
\centering
\includegraphics[width=0.7\linewidth]{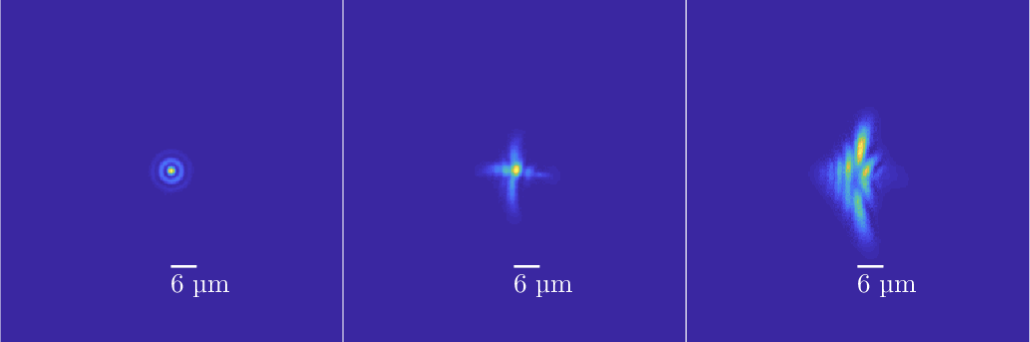}  % fig2 includes two images 
\\
(a) \hspace{0.3\linewidth} (b) \hspace{0.3\linewidth} (c)
\caption{Three measured example PSFs for different image field parameters $(\Delta z, R, \varphi)$. {\bf (a)} (\si{0}, \si{0}, \si{0}), {\bf (b)} (\SI{11.25}{\micro m}, \SI{2.25}{mm}, \si{0}) and {\bf (c)} (\si{0}, \SI{3.00}{mm}, \si{0}). Vertical displacement in (c) due to lens distortion.}
\label{fig:PSFOverview}
\end{figure}

The Point Spread Function (PSF) of an optical system is a highly non-linear function that represents -- in linear system theory -- the transfer function in image space, and is the Fourier transform of the Optical Transfer Function (OTF). In other words, it contains all the information about the performance of the optical system, and the effect of the optical system on a given image can be calculated by convolving the PSF with the input data~\cite{Born1999}. An example measurement of the PSF is shown in Fig.~\ref{fig:PSFOverview}, which belongs to a mass production lens from the automotive industry, and accordingly exhibits strong aberations. 

For certain situations Fourier optics allows to calculate the PSF of a given system by Fourier transforming the aperture. For example the Airy-function is the Fourier transform of the circular step function~\cite{lauterborn2013}. Historically, Zernike developed a set of analytical functions -- the Zernike polynomials -- to describe aberrations in optical systems~\cite{Zernike1934} which are still in wide use. Today, computational imaging is a strong force in the development of novel optical systems, relying heavily on engineering certain PSFs to allow new functions, e.g. extended depth of focus~\cite{Zalevsky2010}, or enhancing digital Fourier microscopy by tailoring asymmetric PSFs~\cite{Wulstein2017}. 

Nonetheless, despite the successes and advances in understanding and using the PSF these numerical and analytical instruments are not well-suited to deal with strong aberrations like the one presented in Fig.~\ref{fig:PSFOverview} as they are present in mass production lenses.
%adequate, i.e. neither computationally efficient nor numerically stable, for optical systems with . In other words, 
Zernike polynomials are ineffective for such lenses, as the aberrations are huge in comparison to diffraction-limited optical systems like telescopes or microscopes. In mass production lenses a large number of Zernike polynomials (50 - 100) need to be taken into account, rendering this process numerically unstable for fitting or modeling of real, measured lenses. All the other examples only work for single, specialized optical systems, but fail for mass production. 

Put simply, what is required  is to be able to measure the PSF of a lens, model that measurement, and then include a numerically efficient version of that model in other optic simulations. A very important example are camera systems for autonomous vehicles, where optical scene simulations are used to both train and test algorithms, like in a computer racing game. So far it is not possible to include the exact optical properties of the real lens used in the product (as given by the PSF) in that simulation. This leaves an important gap in the validation of those cameras systems where the functional and safety limits of the camera system are determined. 

In a previous work~\cite{wittpahl2018} we have demonstrated a novel numerical model for the point spread function (PSF) of an optical system that can efficiently model both experimental measurements and lens design simulations of the PSF. The novelty lies in the portability and the parameterization of this model, which allows to apply this model in basically any conceivable optical simulation scenario. The numerical basis for this model is a non-linear regression of the PSF with an artificial neural network (ANN). In this work we examine two important aspects of this model: the spatial resolution and the accuracy of the model. 

The spatial resolution of the PSF measurement we performed is \SI{0.3}{\micro m}. In a lens design software the spatial resolution is an adjustable parameter. Measurement and simulation can therefore have much higher resolution then the typical pixel size used in current camera sensors, especially those for the automotive industry, where the pixel size is between \SI{3}{\micro m} and \SI{6}{\micro m}. Thus it is not necessary to keep the high resolution, and decreasing the spatial resolution of the PSF measurement by cropping and binning significantly reduces the number of training data and of output neurons. 
%Further it is not obvious what influence this binning has on the accuracy of the training. Therefore we present a systematic study of the resolution of the downsampled model and its influence on the modeling performance. 

Another important influence on the accuracy of the trained ANN is the error metric which is used during training. In our first approach we used the standard mean squared error (MSE), which is used for the presented material. We discuss other metrics and network topologies, and the advantages they promise with regard to the spatial information present in the PSF data. 

This paper is structured as follows. First the model idea itself is presented in Sec.~\ref{sec:ann_regression}. In Sec.~\ref{sec:applications} two important applications are detailed, which are enabled by the optical model. Sec.~\ref{sec:resolution} and \ref{sec:accuracy} then examine the resolution and accuracy, respectively, of the optical model. We close with a discussion in Sec.~\ref{sec:conclusion}. 

%%%%%%%%%%%%%%%%%%%%%%%%%%%%%%%%%%%%%%%%%%%%%%%%%%%%%%%%%%%%%%%%%%%%%%%%%%%%%%
\section{Non-linear regression with artificial neural networks} 
\label{sec:ann_regression}
\begin{figure}[ht]
\centering
\includegraphics[width=0.4\linewidth]{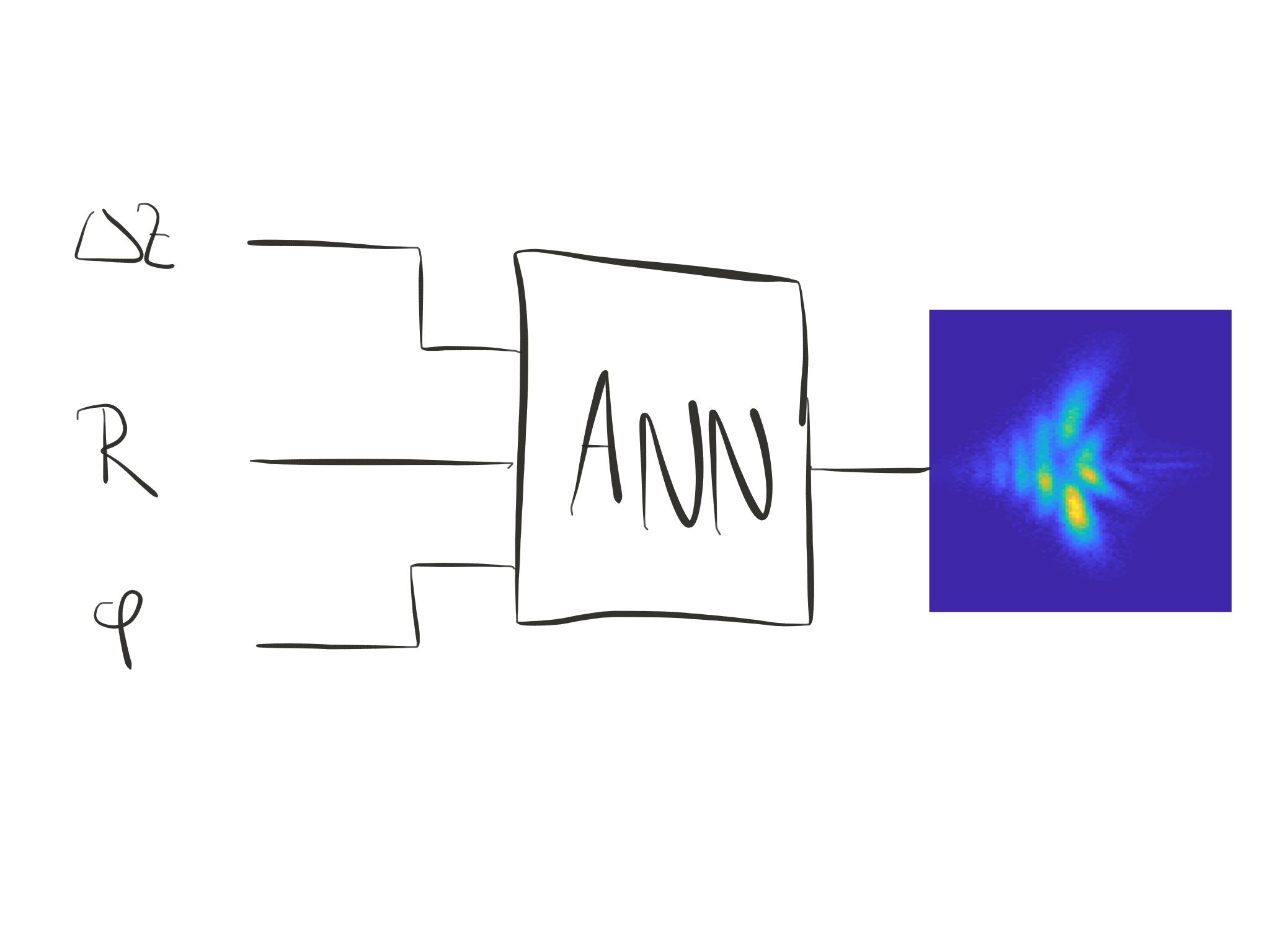}
\caption{Model overview: the ANN takes a number of input parameters (in principle variable) and outputs a PSF. In this report the model parameters are defocus $\Delta z$, image height $R$ and azimuth $\varphi$, according to the measurement parameters.} 
\label{fig:modelOverview}
\end{figure} 
We have a distinctly non-linear function, the PSF, which varies strongly over field and defocus, but nonetheless exhibits strong symmetries and spatial relations. Regression and especially non-linear regression with artificial neural networks for highly asymmetric functions is an established and ongoing field of research\cite{Rosenblatt1962,Rumelhart1986,Belagiannis2015}. The ANN 'learns' the function by training it with an appropriate number of examples. During training both the function training set as well as the according parameters are used as input to the first layer of the ANN. During operation (\emph{inference}) only the input parameters are used, and the ANN evaluates the function at the given parameter position. The goal of our novel model is depicted in Fig.~\ref{fig:modelOverview}. The ANN takes a (limited) number of input parameters and evaluates the output PSF as a function of those parameters. 

The input parameters are in principle variable and depend on the actual simulation goal. E.g. in a tolerance calculation some tolerance measure might be used as input into the ANN. Alternatively, for a rotationally symmetric PSF (from a lens design software in nominal position) the azimuth is not relevant. In this report we use three input parameters to the ANN, defocus $\Delta z$, image height $R$ and azimuth $\varphi$. These parameters accord to the three measurement parameters (see next Sec. Measurement). But the strength of our approach is that it is very flexible in the number and type of inputs, such that it can be used in a broad variety of different simulation scenarios. 

%%%%%%%%%%%%%%%%%%%%%%%%%%%%%%%%%%%%%%%%%%%%%%%%%%%%
\subsection{Measurement data}
\label{sec:measurement}
\begin{figure}[ht]
\centering
\includegraphics[width=0.5\linewidth]{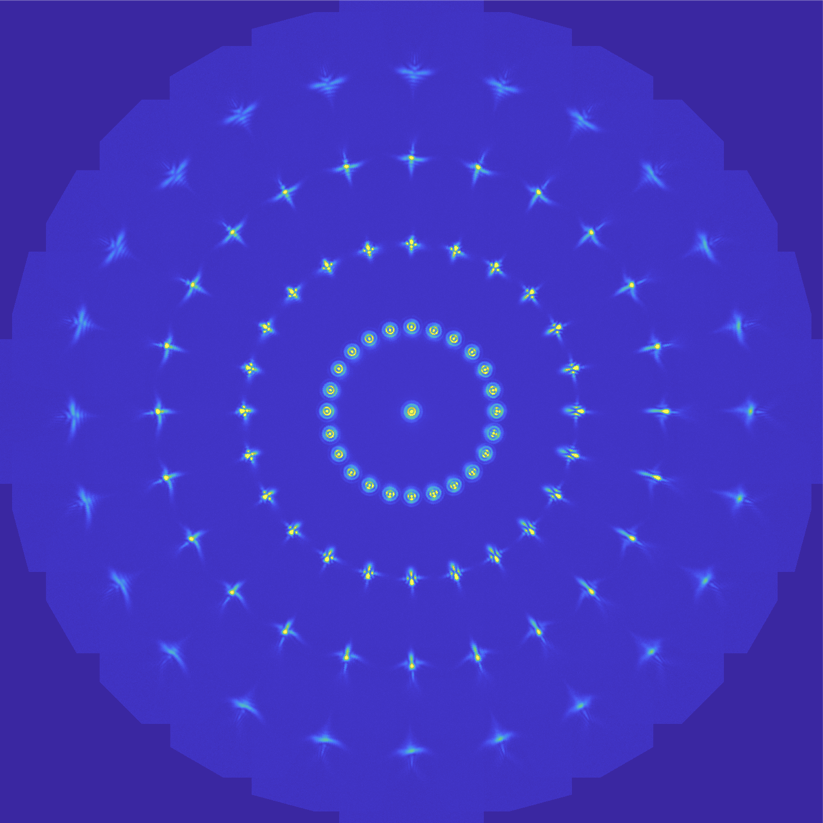}
\caption{Measurement overview with 108 measured PSFs for a single defocus value $\Delta z = \SI{0}{\micro m}$. Radius varied from $R = \SI{-3,00}{\milli\metre}$ to $R = \SI{+3,00}{\milli\metre}$, azimuth varied from \SI{0}{\degree} to \SI{165}{\degree}. Position to scale, PSF enlarged for visibility.} 
\label{fig:MeasurementOverview}
\end{figure} 
The measurement data were taken in collaboration with the company trioptics in Wedel, Germany. The used measurement system was a Trioptics ImageMaster HR. The used setup has an effective pixelsize of $d_\mathrm{pixel, effective} = \SI{0,3070}{\micro\metre}$. Due to the combination of the photopic vision filter and a monochromatic CCD-sensor the measured data does not cover chromatic aberrations. This is left for future work. Overall 27 lenses were measured, based on three different lens designs for automotive camera modules. For this work we selected just a single lens, meaning that the training of the neural network is based solely on the measurement data of one lens specimen. The lens has a focal length of \SI{6}{mm} and a field-of-view of \SI{60}{\degree}. 

The measurement has three parameters: defocus $\Delta z$, image height $R$ and azimuth $\varphi$. The image height was varied from $R = \SI{-3,00}{\milli\metre}$ to $R = \SI{+3,00}{\milli\metre}$, azimuth full circle, and defocus from \SI{-50,0}{\micro m} to \SI{+50,0}{\micro m}. Due to measurement time restrictions the parameter sampling is not evenly distributed. Basically two measurement series per lens were recorded: one with high in-plane resolution for $R$ and $\varphi$ and low defocus resolution $\Delta z$, and one with reduced in-plane resolution and high resolution and larger range for the defocus. The first series resulted in 243 measured PSFs, the second in 972 PSFs. Fig.~\ref{fig:MeasurementOverview} shows the measured PSFs for the used lens in a single plane of defocus, i.e. all the shown PSFs have the same defocus value of $\Delta z = \SI{0}{\micro m}$. Note that the PSFs positions are to scale in Fig.~\ref{fig:MeasurementOverview}, but the actual PSFs have been enlarged to make them visible. Therefore, the outer circle accords to the image height $R = \SI{3,00}{\milli\metre}$, but the real PSFs are much smaller. 

%%%%%%%%%%%%%%%%%%%%%%%%%%%%%%%%%%%%%%%%%%%%%%%%%%%%
\subsection{PSF model example}
The first important step is to downsample the high resolution scans of the PSF: first, if the target image sensor has a pixel size of (e.g.) \SI{3}{\micro m} the high resolution of the measurement (\SI{0,3070}{\micro\metre}) is not necessary. Second, the resolution determines the size of the ANN, the amount of training data and hence the required computing resources. The exact topology of the network and the resolution of the PSF will be detailed in Sec.~\ref{sec:resolution}. For this sutdy we cropped and downsampled the data to a pixel size of \SI{6.5}{\micro m}, resulting in a work resolution of 13x13 for the PSF, mainly restricted due to the limited computing power. 

\begin{figure}[ht]
\centering
\includegraphics[width=\linewidth]{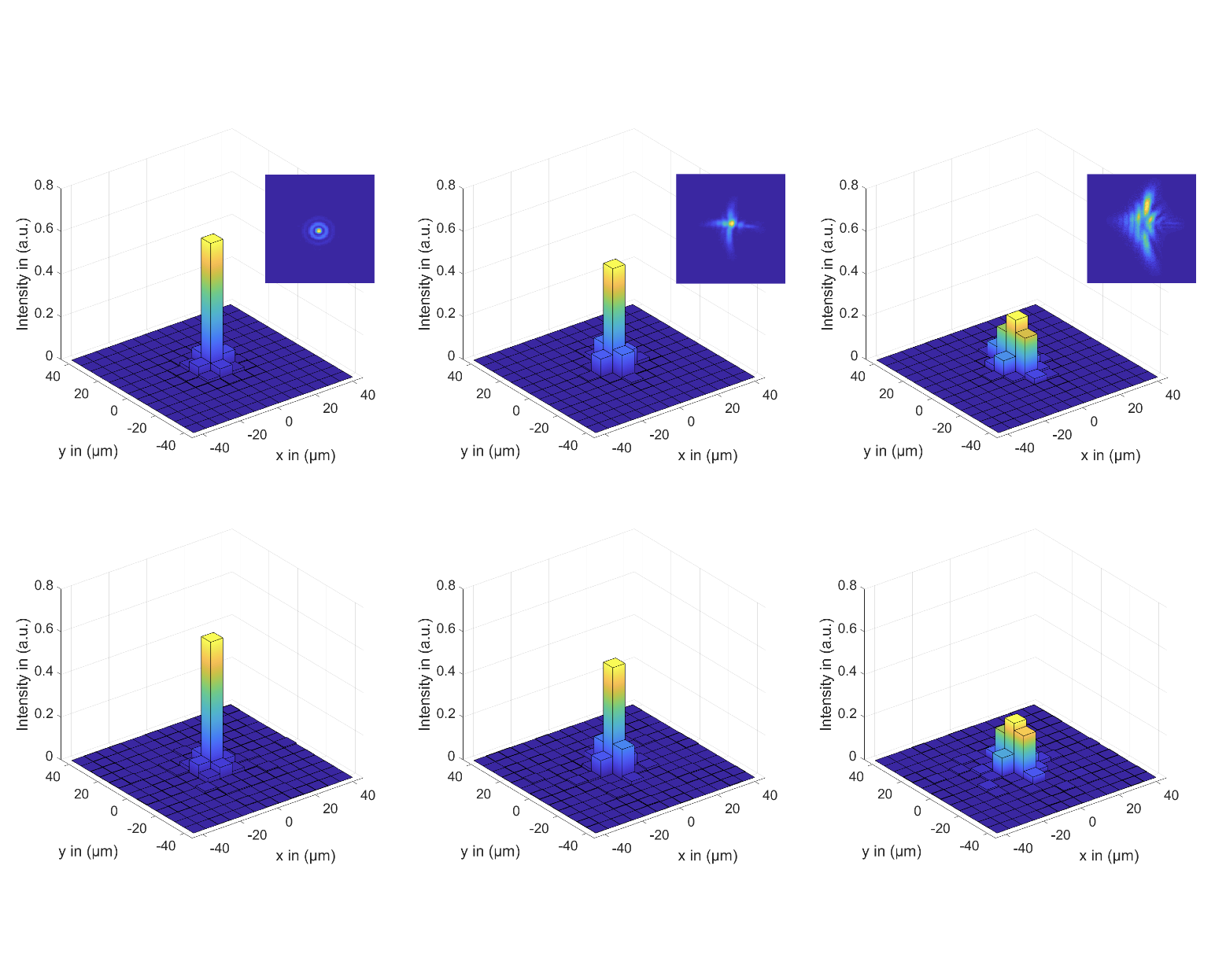}
\caption{Three exemplary modelling predictions for different field situations (cf. text, same as Fig.~\ref{fig:PSFOverview}). Upper row: measurement. Lower row: model prediction. Inset: high resolution PSF.} 
\label{fig:predictionExample}
\end{figure} 
Fig.~\ref{fig:predictionExample} shows an example of the prediction process for a given lens sample, where the PSF is predicted at a position that had been measured.  The three columns a, b and c represent three different field positions $(\Delta z, R, \varphi)$, the same as in Fig.~\ref{fig:PSFOverview}: (a) (0, 0, 0),  (b) (\SI{11.25}{\micro m}, \SI{2.25}{mm}, 0) and  (c) (0, \SI{3.00}{mm}, 0). The upper row shows the actual measurement, the lower row the output from the ANN for the given input parameters. The inset displays the high resolution measurement image of the respective PSF. The ANN can model the distinctly varying spatial distribution of the PSF, especially the changing quality is modeled faithfully. 

\begin{figure}[ht]
\centering
\includegraphics[width=\linewidth]{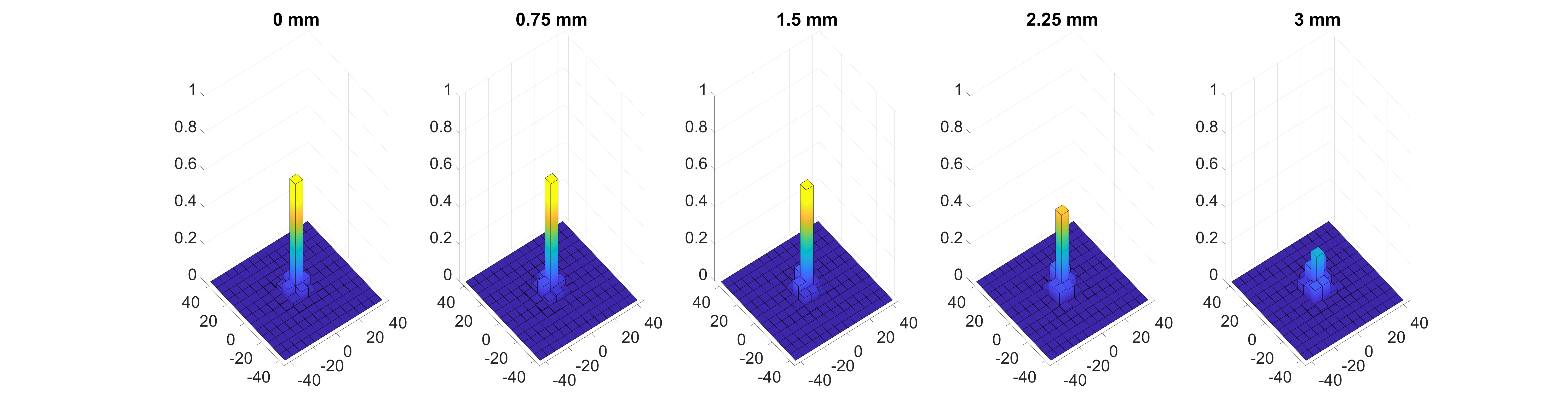}
\caption{Three exemplary modelling predictions for different field situations (cf. text, same as Fig.~\ref{fig:PSFOverview}). Upper row: measurement. Lower row: model prediction. Inset: high resolution PSF.} 
\label{fig:psf_no_measurement}
\end{figure}

As a second simple test we predicted the PSF at positions that had not been measured at all. For the PSF series shown in Fig.~\ref{fig:psf_no_measurement} we turned the azimuth to \SI{37.5}{\degree}, right between the measurements at \SI{30}{\degree} and \SI{45}{\degree}, and increased the field in five steps from zero to maximum (= \SI{3.00}{mm}). The model prediction accords to expectation. 

In summary, the PSF of the given lens can now be predicted for any position within the paramter space of the measurement. The next chapter discusses applying the PSF as a convolution filter for different applications. 

%As an example, in theory a typical imaging lens has rotational symmetry, which in real samples is slightly broken by production tolerances. At the same time these aberrations show a strong correlation with the three image parameters defocus, field height and azimuth. 

%%%%%%%%%%%%%%%%%%%%%%%%%%%%%%%%%%%%%%%%%%%%%%%%%%%%%%%%%%%%%%%%%%%%%%%%%%%%%%
\section{Applications}
\label{sec:applications}
The following section provides two example applications that would currently not be possible without the new model: re-using existing recorded data, and correct imaging of object-space scene depth. 

%%%%%%%%%%%%%%%%%%%%%%%%%%%%%%%%%%%%%%%%%%%%%%%%%%%%
\subsection{Re-use of existing images and video sequences}
Training autonomous vehicles requires lots of driving sequences in all situations\cite{zhao2016}. Typically a simulation environment (software-in-the-loop, SiL) accompanies real-world test drives to systematically vary environmental parameters. In a next step the actual camera system is fed with simulated or pre-recorded data in a hardware-in-the-loop environment (HiL). Unfortunately, as soon as a single optical property changes the whole recorded scene catalogue becomes obsolete, because the data does not reflect the new optical system. Therefore, the goal of a validating SiL or HiL setup is to simulate or modify scenarios to look like they were taken under different circumstances. For example, the temperature expansion of the plastic holder may lead to a defocus $\Delta z$, giving a blurred image. Therefore the image data -- either simulated or already recorded data -- needs to be modified to reflect these circumstances. For the optical question at hand this means that the existing images need to be convolved with the appropriate PSF as a transfer function. 

The original scenes were recorded with a real camera with its own tolerances and optical properties. Therefore the first step is to de-deconvolve the original image or video sequence with the PSFs of the recording lens. This is also one step that so far has not been possible. Instead, several research approaches exist that estimate the PSF (or rather, typically, the MTF) of the optical system from the content of the image or video sequence (blind deconvolution~\cite{Lam2000}, non-blind deconvolution~\cite{Sun2014}). After the deconvolution of the old lens the new lens is then convolved onto the image(s) as a transfer function. After this the image looks like it were taken with the new lens. A possible application therefore would be to record a driving sequence with a well-aligned and sharp camera system, and later on to simulate a defocussing by convolving the sequence with an increasing defocus value $\Delta z$. 

\begin{figure}[ht]
\centering
\includegraphics[width=0.8\linewidth]{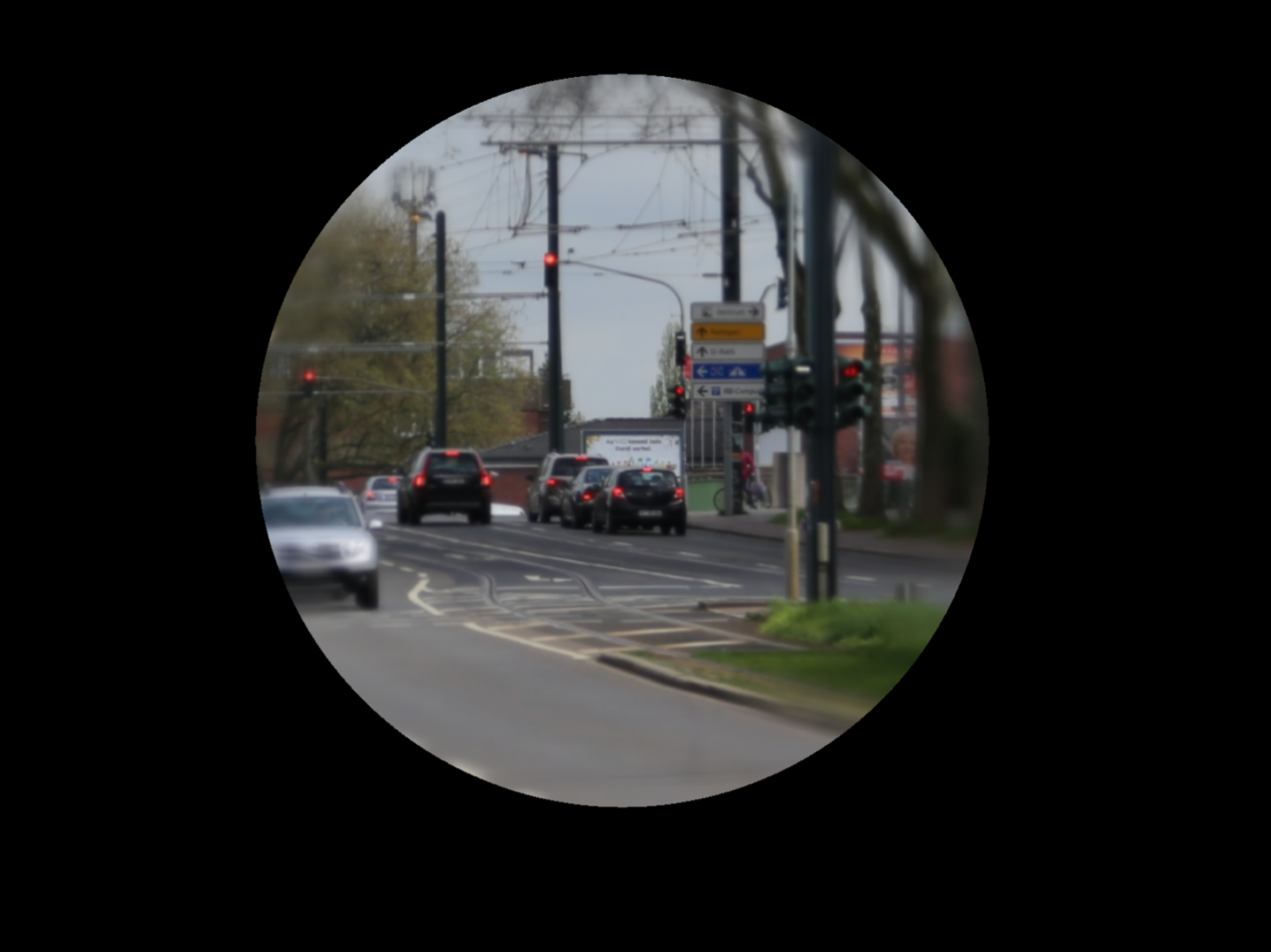}
\caption{Resulting degraded image after applying the spatially variant PSF, with a defocus value of $\Delta z = 0$.}
\label{fig:result}
\end{figure}
An example image produced by this process is shown in Fig.~\ref{fig:result}. There, an image taken with a (very) high quality consumer DSLR with appropriate lens.
For completeness we also applied a standard blind deconvolution (aberration-limited) to minimize the influence of the original lens -- for accuracy this probably wasn't necessary considering the huge difference in optical quality between a state-of-the-art consumer lens and a series production automotive lens, but was also included in the algorithm as a place holder function for a more rigorous examination. Because the physical size of the original image is larger then the maximum radius of our measurement ($R_\text{max} = \SI{3}{mm}$) the performance of the measured lens strongly declines toward the edge of the image circle. In a real application the size of the imager would be selected as a rectangle distinctly within the circle, hence the very strong blurring at the edges would not be visible. 

%%%%%%%%%%%%%%%%%%%%%%%%%%%%%%%%%%%%%%%%%%%%%%%%%%%%
\subsection{Object space depth simulation}
For the simulation shown in Fig.~\ref{fig:result} the defocus value is $\Delta z = 0$ for the whole image, i.e. the depth information (or an inclusion of the field curvature) is not present. This leads to the second application that so far has not been possible in the proposed way: simulate object space depth for a real lens. Object space depth also leads to a defocus, as the image point is formed at a different position then the focal length. This is demonstrated by the simple formula used for geometric optics of image formation:
\begin{equation}
1/f = 1/o + 1/i,
\end{equation}
where $f$, $o$ and $i$ are focal length, object and image distance, respectively. The actual defocus value $\Delta z$ also depends of course on the aligned focus position (in other words the hyperfocal distance in object space). Of course optical design software (like \emph{Code V} or \emph{OpticStudio}) is easily capable of simulating object space depth, but only one object distance at a time. The difference to our model therefore lies in the huge difference in numerical effectiveness which is required to actually reuse terabytes of pre-recorded scene data. 

\begin{figure}[ht]
\centering
\includegraphics[width=0.6\linewidth]{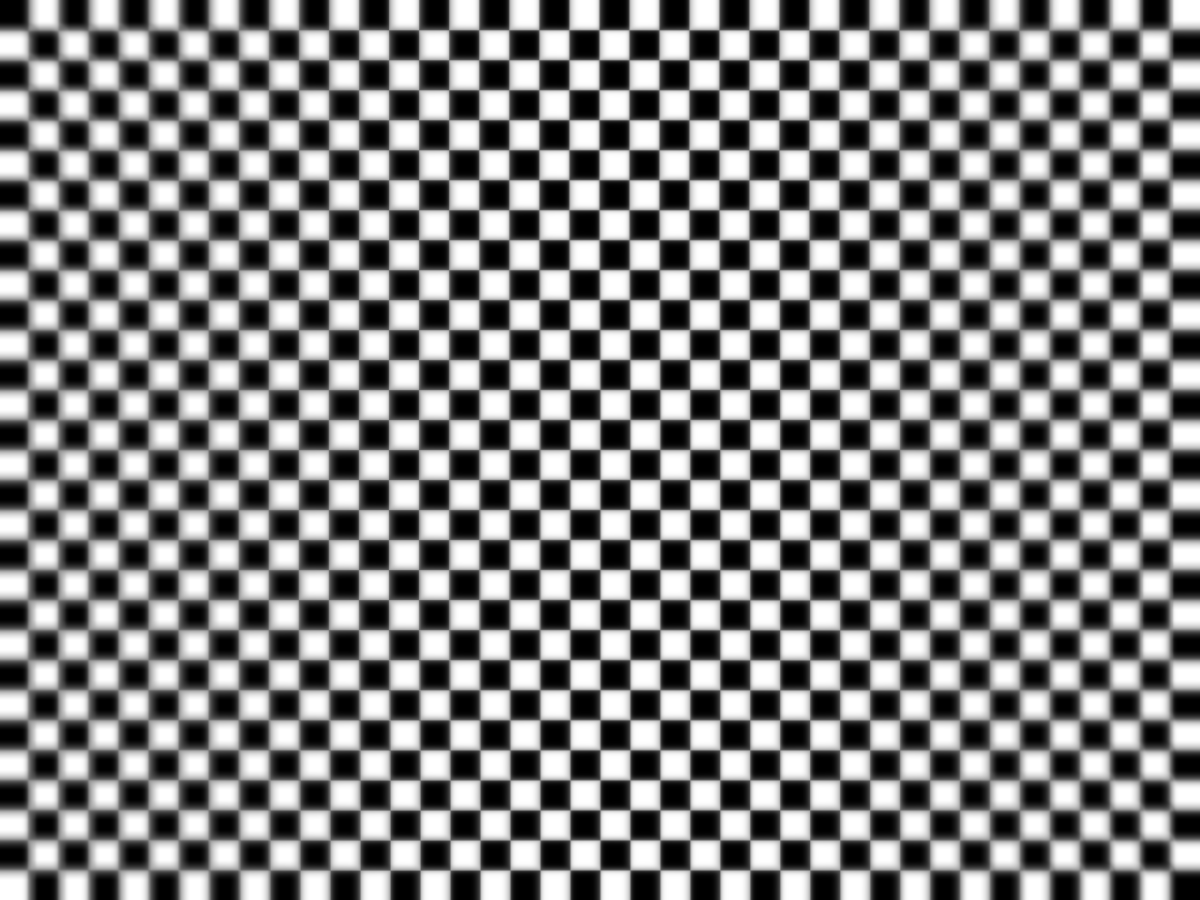}
\caption{Linear depth gradient applied to a checkerboard. Please note the slight differences between the left and right edge of the image, resulting from positive and negative $\Delta z$ values.}
\label{fig:depthCheckerboard}	
\end{figure}
To demonstrate this feature we have applied the model with a linear depth gradient to a simple checkerboard (see Fig.~\ref{fig:depthCheckerboard}). In the image the value for $\Delta z$ is varied from left to right from positive maximum measurement value (here: \SI{+50}{\micro m}) to negative maximum (\SI{-50}{\micro m}).  This setup corresponds roughly to a checkerboard in object space that is strongly tilted with respect to the optical axis of the camera system. Note also the slight difference in aberrations on the left and right hand side of the image, as a positive defocus value accords to a different PSF then the same defocus value in negative direction. In other words, not only is our model spatially variant over image height, but also realistically variant in $z$-direction. Whats more, the model is neither rotationally symmetric nor in $z$-direction, reflecting real aberrations of a mass produced lens.   

%%%%%%%%%%%%%%%%%%%%%%%%%%%%%%%%%%%%%%%%%%%%%%%%%%%%%%%%%%%%%%%%%%%%%%%%%%%%%%
\section{Resolution}
\label{sec:resolution}
This section clarifies the definition of the different resolutions that are used in the model and its application. First, both the PSF data and the image data have spatial and intensity resolutions. Second, this has an impact on the topology of the artificial neural network. 

%%%%%%%%%%%%%%%%%%%%%%%%%%%%%%%%%%%%%%%%%%%%%%%%%%%%
\subsection{PSF and image resolution}
The model and the applications have two different resolutions: first the resolution of the measurement or simulation, and second the application resolution. In our case the measurement resolution is determined by the hardware of the ImageMaster HR, which consists of a microscope and a measurement sensor. The resulting resolution is $d_\mathrm{pixel, effective} = \SI{0,3070}{\micro\metre}$ (cf. Sec.~\ref{sec:measurement}). The simulation resolution using a lens design software is a free parameter. In both cases the PSF is recorded as a 2d-matrix of intensity values. 

This PSF is used as a convolution filter kernel in an application (see Sec.~\ref{sec:applications}). Here the resolution is determined by the actual pixel size of the used camera. As an example, a simple ADAS camera typically has VGA resolution and a pixel size of approximately \SI{6}{\micro m}, almost 20 times more then the resolution of the PSF. As both the PSF data and the application image(s) are matrices of discrete data the scales need to be adapted. This scale adaption can be done both in image and frequency space, i.e. after Fourier transforming both the PSF and the target image. Here we chose to scale the PSF in image space and apply the convolution as a convolution in image space, not as a multiplication in frequency space. 

The original PSF measurement data is approx. 1400 by 1100 pixel. We first crop the data to 256 by 256 pixel. Then, the image is subsampled and linearly interpolated to achieve the scale adaption. The result is a subsamples PSF of size 13 x 13, which accords to a target image resolution of \SI{6.5}{\micro m}. An example is shown in Fig.~\ref{fig:psf_no_measurement}. Now this matrix may be directly applied as a convolution kernel on the target image. 

So far we have not talked about the bit depth resolution (i.e. dynamic range) which needs to be adapted as well. The PSF data has \SI{12}{bit} resolution, and the target image from the high-quality DSLR has \SI{16}{bit} resolution. This is currently solved by casting the PSF data to floating point and normalizing the values such that the volume of the PSF is one. This accords to energy conservation with respect to the input energy. This is technically not correct, as the reduced PSF amplitude is a measure of the transmitted light (including both vignetting and $\cos^4$-fall-off), but is left for future work.

Further details are left for future work. The bit depth resolution becomes tricky when the image sensor uses HDR images, which currently is the standard in ADAS cameras. Further, color is not handled at all. Here the spectral resolution in the ImageMaster HR, the CFA of the imager and the demosaicing process all need to be taken into account. Lastly, the distortion of the lens is currently also ignored. This shows as a decentering of the PSF in the measurement images. The decentering value itself could be used to construct a rectification function, but in this first step the PSF distributions are centered on the centroid before cropping to 256 by 256 pixels. 

%%%%%%%%%%%%%%%%%%%%%%%%%%%%%%%%%%%%%%%%%%%%%%%%%%%%
\subsection{Resolution and topology of the ANN}
\begin{figure}[h]
\centering
\includegraphics[width=0.6\textwidth]{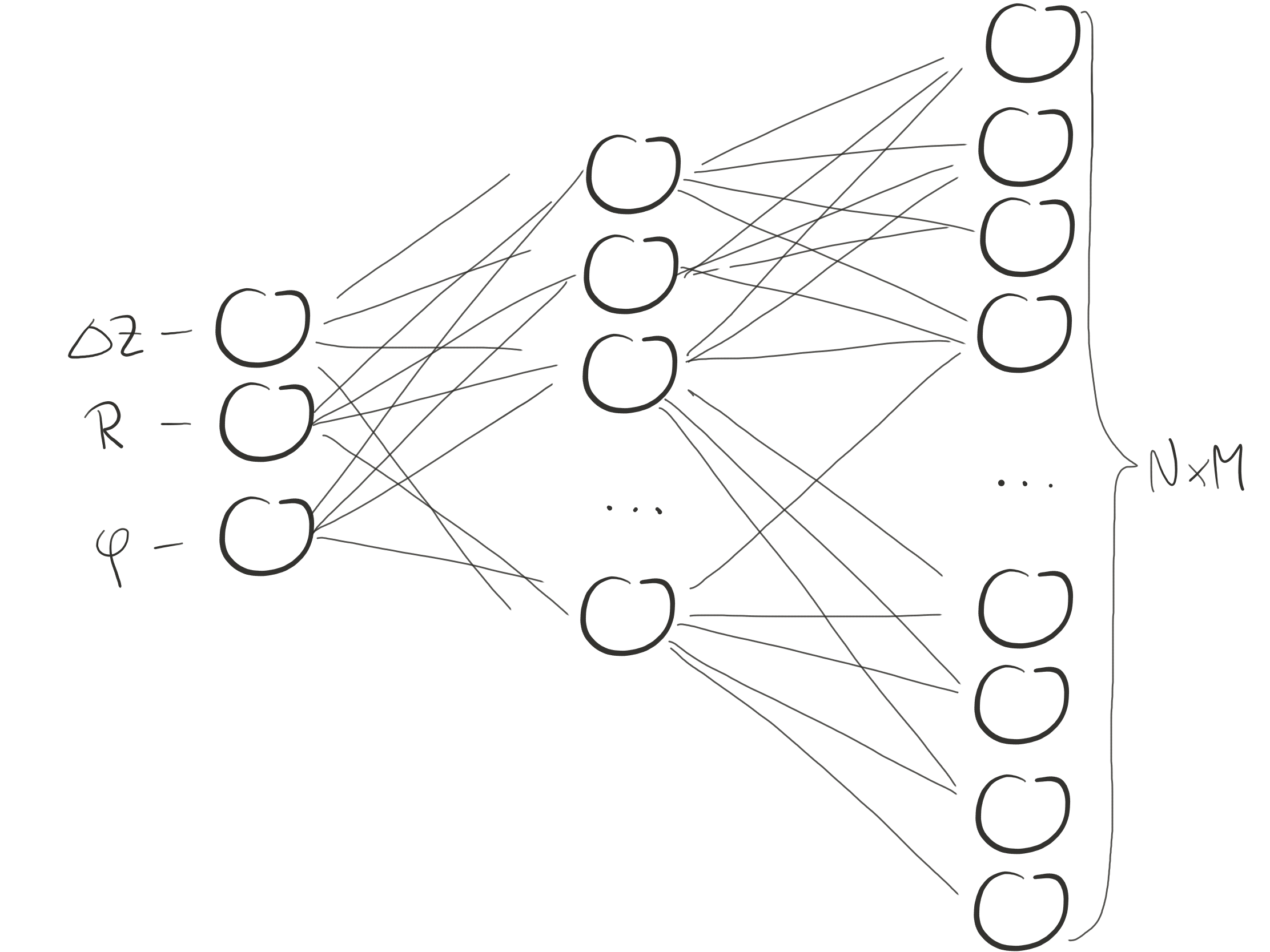}
\caption{Topology of the used ANN. The input layer has three neurons corresponding to the measurement parameters defocus $\Delta z$, image field $R$ and azimuth $\varphi$. We use only one hidden layer which is varied, but has approximately 70 - 100 neurons. The number of output neurons equals the number of pixels in the predicted PSF, in this case 169. The net is feed-forward connected.}
\label{fig:topology}
\end{figure}

The artificial neural network performs a non-linear regression of the PSF data, in other words: the output of the ANN is the PSF itself. Therefore, the number of neurons in the output layer of the ANN is equal to the number of pixels in the PSF prediction as seen in Fig.~\ref{fig:topology}. If we train the ANN on the high resolution measurements directly this would imply 1540000 neurons at a resolution of 1400 by 1100 pixels. This is not only numerically cumbersome, but physically not necessary. As described in the previous section the application requires a much lower spatial resolution of the PSF, hence we are able to directly train the ANN with the downsampled data, reducing the number of output neurons to 169 (13 by 13) at \SI{6.5}{\micro m} application pixel size. 

From Fig.~\ref{fig:topology} it is apparent that the number of output neurons directly influences the amount of computational requirements (memory, calculation time), so from that perspective the lower would be better. Furthermore, we predict the value of 169 neurons (in the shown example) based on the value of three neurons in the input layer. From an information theoretical point of view it is not obvious that this creation of new information is numerically stable for any output resolution, i.e. there may be a limit to the spatial resolution of the predicted PSF based only on three input parameters. 

On the other hand, the measurement data itself shows the strong correlation the PSFs have to the input parameters. As an example, select any ring of PSFs in Fig.~\ref{fig:MeasurementOverview}. For the same value of $R$ all the PSFs look basically the same\footnote{If this were simulation data with rotational symmetry they would be exactly the same.}, and the variation due to $R$ is much stronger than the variation due to the azimuth $\varphi$. In other words, while we may have changed only one parameter $R$ it is obvious to the human eye (and mind) that the resulting structure is predictable -- which is exactly why an artificial neural network is predestined to perform the non-linear regression which we propose. 

The questions remains how far the spatial resolution of the PSF can be increased, which so far we discussed theoretically. From a practical point of view there are two more aspects to consider. For one, the predicted PSF shall be used as a convolution kernel, and execution speed again prefers the smallest possible resolution -- which in the discussed cases is the pixel size of the application imager. 

But secondly, the whole model is very interesting for the optical system design of complete camera systems. During design stage the imager is not fixed, but selected from a list of possible candidates with a complete set of different requirements. In an optical simulation that covers the complete camera system it would be a great benefit if the imager were just an exchangeable part. In this scenario the scale adaption of the PSF from the measurement to the application should be following the simulation of the light distribution, in other words: applying the PSF as a transfer function should be high resolution, and these results then need to be cast to the right scale depending on the imager pixel size. 

Summarizing, the spatial resolution of the PSF should be as small as possible to minimize calculation cost for both the training of the neural network and the application as a convolution kernel. On the other hand the resolution should be as high as possible to retain some flexibility in a camera simulation, such that different imagers may be tested with the same optical lens simulation. 

%%%%%%%%%%%%%%%%%%%%%%%%%%%%%%%%%%%%%%%%%%%%%%%%%%%%%%%%%%%%%%%%%%%%%%%%%%%%%%
\section{Accuracy}
\label{sec:accuracy}
Accuracy in this context means how close the prediction models the actual measurement (simulation). 
%%%%%%%%%%%%%%%%%%%%%%%%%%%%%%%%%%%%%%%%%%%%%%%%%%%%
%\subsection{Accuracy using Mean Square Error}
The first obvious metric is mean square error (MSE), i.e.
\begin{equation}
\text{MSE} = \sqrt{\sum_i \left (x^\text{model}_i - x^\text{measured}_i \right)^2},
\end{equation}
where $x^\text{model}_i$ and $x^\text{measured}_i$ are the predicted and the measured value, respectively, for the $i$th pixel of the PSF. This is the metric the results in this work are based on. For the training of the ANN the square matrix PSF data was converted into a linear array of pixel values. The ANN was then trained with different training algorithms, and the quality function for the training was just the MSE. The MSE distance metric yields satisfactory agreement between the measurement and the model, as demonstrated above (cf. Fig.~\ref{fig:predictionExample} and Fig.~\ref{fig:psf_no_measurement}). 

\begin{figure}[h]
\centering
\includegraphics[width=0.6\textwidth]{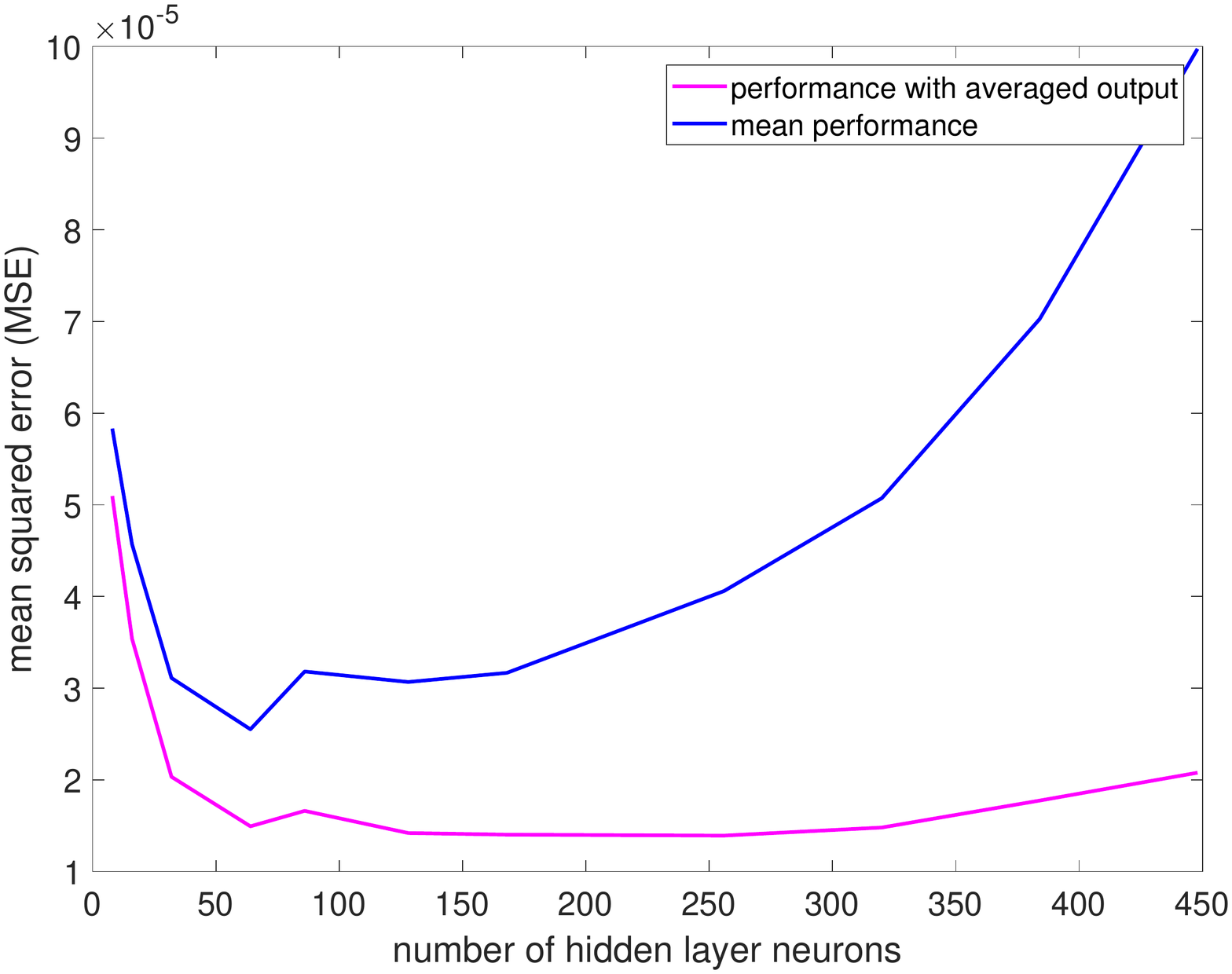}
\caption{Variation of the number of hidden neurons between 8 and 448. For each configuration the training was started a 100 times with a different randomized bias. Shown are the mean performance and the averaged output performance (cf. text). }
\label{fig:numHiddenNeurons}
\end{figure}
Using MSE as our metric we've systematically varied the number of hidden neurons to find the optimal topology for the ANN. Fig.~\ref{fig:numHiddenNeurons} depicts the results of this variation, where the number of hidden neurons was changed from 8 to 448. For each configuration the training was started a 100 times with a different randomized bias. Shown are two graphs: the mean performance (i.e. the average of the 100 resulting MSEs) and the performance for the averaged output (i.e. averaging the 100 values for each output neuron, then taking the MSE). The mean performance shows the typical behavior of a ANN: first the quality metric improves a lot (= smaller MSE), finds a plateau-like optimum -- here at approximately 50 - 100 neurons -- and then the MSE increases again (= worse performance). This increase in MSE for higher number of neurons is typical for ANNs: the net starts to memorize the data and is less capable of generalizing to data it hasn't seen before. 

%%%%%%%%%%%%%%%%%%%%%%%%%%%%%%%%%%%%%%%%%%%%%%%%%%%%%
%\subsection{Possible improvements}
While the MSE as a metric yields satisfactory results, from an physical (optical) point of view ignoring the spatial information present in the 2d matrix of the PSF seems an omission. We currently see two ways to take the spatial information of the PSF into account. First, a different metric is required which focuses on the \emph{similarity} of the measured and the modeled PSF. E.g. both functions are concentrated around the center of the matrix, and hence edge pixels should not contribute as much. This is an ongoing and active field of research, see e.g.~\cite{Dziugys2015} and references therein. 

Secondly, the whole structure of the ANN may be adapted to reflect the spatial properties of the data. Convolutional Neural Networks (CNN) are an established tool to work with 2d image data. The convolutional part of the neural network requires a certain topology of the neural network, where the spatial relation of the data is already reflected in the number of neurons and the type of connections. Put simply, a convolution kernel is a small 2d structure of filter weights. Apparently, a neural network can reflect this structure by the number of connections and its weights. Currently, CNNs are used to classify image data into categories, i.e. a $M\times N$ image matrix is mapped onto a number of output neurons, where each neuron represents the probability of a given category. This is approximately the inverse of our model (cf. Fig.~\ref{fig:topology}), so the application of CNNs to our non-linear regression on the output side is not apparent. Nonetheless, we are currently pursuing both avenues -- spatially aware metric and CNNs -- to improve the accuracy of our PSF model. 

Summarizing, the accuracy of the ANN depends on both the metric used during the learning phase, as well as the topology of the neural network itself. Encoding spatial information in an ANN is a highly dynamic and evolving field of research, and fruitful new discoveries are to be expected that can then be applied to improve the accuracy of our PSF prediction. 

\section{Conclusion}
\label{sec:conclusion}
We have presented our novel numerical model for the PSF of an optical system, with the goal of using measured data in a simulation scenario. New simulation applications are now possible, like re-using existing drive scene catalogues and including object-space depth information in a numerically efficient manner. This is especially useful in the context of autonomous driving, where the validation of a camera system needs to precisely determine the functional limits of the system, which so far has been difficult due to the lack of an appropriate optical model. 

Both the resolution and the accuracy of the optical model are determined by the topology of the ANN. The right resolution is a trade-off between application resolution (both spatial and dynamic range) and computational requirements. Finally, there are many different avenues to pursue in optimizing the accuracy of the prediction, based on the ongoing research in the domain of 2d-processing artifical neural networks. 

\section*{Acknowledgement}
We gratefully acknowledge the support from the company Trioptics, which allowed us to measure sample lenses in their own development lab with state-of-the art lens measuring equipment. 
% References
%\bibliography{report} % bibliography data in report.bib
%\bibliographystyle{spiebib} % makes bibtex use spiebib.bst
%%%%%%%%%%%%%%%%%%%%%%%%%%%%%%%%%%%%%%%%%%%%%%%%%%%%%%%%%%%%%%%%%%%%%%%%%%%%%%%
% Bib
%%%%%%%%%%%%%%%%%%%%%%%%%%%%%%%%%%%%%%%%%%%%%%%%%%%%%%%%%%%%%%%%%%%%%%%%%%%%%%%
\clearpage
\small
\bibliographystyle{spiebib}
\bibliography{Literature}

\end{document}